# Reflection On Reflection In Design Study

Jason Dykes, City, University of London
Miriah Meyer, University of Utah

## Abstract

Visualization design study research methodologies emphasize the need for reflection to generate knowledge. And yet, there is very little guidance in the literature specifying what reflection in the context of design studies actually involves. We initiated a community discussion on this topic through a panel at the 2017 IEEE VIS Conference -- this report documents the panel discussion. We analyze the panel content through the lense of our own reflective experiences and propose several priorities for ongoing thinking on reflection in applied visualization research.

## Introduction

In 2012 Sedlmair et al. proposed a design study methodology, laying out a call to the visualization community to take a more structured and methodical approach to design-focused applied visualization research [1]. While some guidance in the methodology is specific to design studies, there are many ideas that cross-cut applied visualization research more generally. One of these is the role of reflection in generating and validating knowledge from highly situated research settings.

The paper notes that "*reflection is where research emerges from engineering*", and provides this guidance for reporting on design study research: "*a careful selection of decisions made, and their justification, is imperative for narrating a compelling story about a design study and are worth discussing as part of the reflections on lessons learned*." Best practice according to the paper might involve "*reflecting on lessons learned from the specific situation of study in order to derive new or refined general guidelines [, which] typically requires an iterative process of thinking and writing*." This limited guidance is the extent of existing advice in the visualization literature on the role of reflection, and what it might entail for applied visualization research.

In our own research groups we use reflection throughout the process of designing new visualization techniques and systems, as well as when we try to understand the broader implications of the applied research that we do. The specifics of our reflective practices, however, are not always the same. Furthermore, we don't actually know how we should be reflecting or even how we *could* be reflecting — the limited discussion in our literature is not enough. What works, what doesn't, and how certain are we about any of this?

These questions led us to look at research communities beyond visualization that engage heavily in reflective practices, including education, management, and healthcare. The literature from these fields present a range of definitions for reflection. For example, in the seminal work on the topic, Schon defines reflection as "*the practice by which professionals become aware of their implicit knowledge base and learn from their experience*" [2]. In McGill and Brockbank's pragmatic guide, they provide this definition: "*a process by which experience is brought into consideration … to achieve meaning and the capacity to look at things as potentially other than they appear*" [3]. More recently, Bolton characterized reflection as "*paying critical attention to the practical values and theories which inform everyday actions, by examining practice reflectively and reflexively… leading to developmental insight*" [4].

These definitions predominantly focus on reflection in practice — do they apply to research? And if so, how? More specifically, how do they apply to *visualization* research? We believe that the visualization community is lacking consensus as to the role of reflection in our research practices, as well as an articulation of standards of good practice for reflection within design studies, and applied visualization research more broadly. This leads us to ask: how do we use reflection to make implicit knowledge explicit, to interpret what we observe in applied contexts critically and authentically, and to use experience and multiple perspectives to derive reliable knowledge from the people, software, and contexts that we study in our visualization research?

To reflect on the role of reflection in design studies we organized a panel on the topic at the 2017 IEEE VIS conference, the premier venue for visualization research. The experience of the panelists spans the gamut of visualization approaches: qualitative analysis, controlled studies, technique and systems design, methodology, and design studies. Some of the panelists conduct design studies, some consume them, and one of them established their role in the visualization community. The audience consisted of a cross section of conference attendees with expertise across visualization, and who participated in the conversation through traditional questions and answers as well as through live polls conducted using web-based technology.

In this report we describe the content of the panel session and provide a short, reflective summary of key themes we identified in the content. The content was captured through an audio recording of the panel, notes taken during the session and a spreadsheet that logged audience responses to key questions raised through the live polls — this content is publicly available via links in this paper. Based on this summary and our own reflective experiences, we propose several priorities for ongoing thinking on reflection in applied visualization research.

# Panelists

**REMCO CHANG**

Associate Professor in the Computer Science Department at Tufts University, Remco received his PhD in Computer Science from UNC Charlotte in 2009. Prior to his PhD, he worked for Boeing where he developed real-time flight tracking and visualization software. His current research interests include visual analytics, information visualization, human-computer interaction, and databases.
http://www.cs.tufts.edu/~remco/

**NATHALIE HENRY RICHE**
A researcher in the Extended Perception Interaction Cognition group at Microsoft Research, Nathalie holds a Ph.D. in computer science from the University of Paris-Sud/Inria and the University of Sydney. Her research interests span human-computer interaction with a focus on data and information visualization. Nathalie is a firm believer in user-centered and participatory design methods and creates novel interactive visualizations for exploring, thinking and communicating with data with people who need and use these systems.
https://www.microsoft.com/en-us/research/people/nath/

**UTA HINRICHS**
Lecturer at the Human-Computer Interaction Group at the University of St Andrews, Uta holds a PhD in Computer Science with specialization in Computational Media Design from the University of Calgary. Uta's research, driven by interdisciplinary collaborations, focuses on information exploration and analysis as part of professional activities and everyday life. She combines design explorations with qualitative research methods to investigate in-situ how visual information exploration tools are being used.
http://www.utahinrichs.de

**PETRA ISENBERG**
A research scientist at INRIA, Saclay, France, Petra holds a PhD from the University of Calgary. Her main research areas are information visualization and visual analytics with a focus on collaborative work scenarios, interaction, and evaluation. Petra is interested in exploring how people can most effectively work together when analyzing large and complex data sets on novel display technology such as small touch-screens, wall displays, and tabletops.
https://petra.isenberg.cc/

**HEIDI LAM**
A research scientist at Tableau Software, Heidi received a PhD from the University of British Columbia. Her current research focuses on understanding how visualization can be used to support data analysis, as well as finding new methodologies to achieve that understanding. A tool-builder at heart, Heidi hopes that such understanding can be applied to visualization tools to enable a wider set of people to better understand our complex and interesting world.
https://research.tableau.com/user/heidi-lam

**TAMARA MUNZNER**
Professor of Computer Science at the University of British Columbia, Tamara holds a PhD from Stanford. She has worked on problem-driven visualization in many domains including genomics, computational linguistics, web log analysis, and journalism. Beyond coining the term *design studies*, Tamara is also known for her technique interests in graph drawing and dimensionality reduction, as well as her evaluation interests including controlled experiments in a laboratory setting and qualitative studies in the field.
http://www.cs.ubc.ca/~tmm/

# Methods

As co-organizers, we developed initial ideas for the panel content based on our own interest in qualitative research methods and applied visualization research. We focused on topics that we believe to be poorly understood and ill-defined by the visualization research community, and found reflective practice to be a core thread. Our selection of panelists was then based on their expertise that cross-cuts both reflective practice and applied visualization research.

Once the panelists were confirmed we sent each an initial set of questions to guide their individual reflection on the panel topic, and asked for short responses in return. The questions were:

1. When and how do you reflect when conducting a design study?
2. How do you record the process and results of any reflection during a design study?
3. How might we make contributions generated through reflection in design studies more trustworthy and useful?
4. How and when does cross-study reflection occur? Do you learn from, or transfer knowledge to other studies beyond your own experience?
5. Has this reflection on your own practice resulted in any further thoughts or ideas about reflection in design studies?

We compiled the panelists' responses and sent them around to the group. We asked that each panelist consider all other responses when developing position statements for the panel session. The position statements were tightly constrained to ensure that messages were clear and that there was plenty of room for discussion: 1 slide, presented in 4 minutes. Additionally, we asked that panelists structure their statements around the following questions:

1. What do you think reflection is?
2. Do you think reflection is important in visualization research? If so, *how* and *when*? If not, *why not*?
3. And then, depending on your interest and experience, address either…
   - Where and how does reflection occur in your design study research?
   - Provide some examples of the results of reflection that you have used in your research or design.

At the IEEE VIS conference, hosted in Phoenix, Arizona (October 2017), we invited the panelists to lunch one day prior to the panel session to discuss the panel logistics — all accepted our invitation. We deliberately kept the conversation away from any discussion of the panel topic itself but did give the panelists opportunities to get to know each other a little and ask questions about panel format and expectations. As organizers we also reviewed the panelists' position statements at the conference and developed a targeted question for each to ask during the session.

The panel session followed a traditional format: it began with an opening statement by the organizers, followed by position statements from each panelist, and concluded with an interactive question and answer phase in which the audience raised issues for discussion. Additionally, we included web-based polling technology (www.polleverywhere.com) to collect ideas and questions from the audience throughout the session. The live polling results were displayed and discussed during the question and answer phase.

The polls initially asked the audience for two pieces of information:

1. What is your area of expertise?
2. When do you reflect?

Following the question and answer phase we then polled for ideas about reflection stimulated by the discussion and thinking that had occurred in the session. These were solicited in three categories. We gave the panelists and audience ten minutes to log their contributions through a poll:

- **best practice:** what ways of reflecting have worked well for you?
- **possible pitfalls:** what has, or could go wrong?
- **open questions:** what else do we need to know about reflection?

After the session we collected two sources of raw data from the panel — an audio recording of the session and the poll results. One of the organizers cleaned the poll results and analyzed the data using thematic analysis [5], and the other created a summarized transcript from the audio recording and notes. We individually listened to the audio recording of the session and reflected on the content before coming together to: develop a coarse summary of the data content; draw some initial conclusions; and identify areas for further work and analysis.

# Data

The first data source of the panel content are responses made to three polls conducted during the session. The responses are provided in a spreadsheet that contains a worksheet for each poll, with each response tagged with its submission time: http://j.mp/googleSheet_IEEEVIS17_reflection. Figures 1 and 2 present overviews of the responses to the first two poll questions: 1) What is your area of expertise?; and 2) When do you reflect?

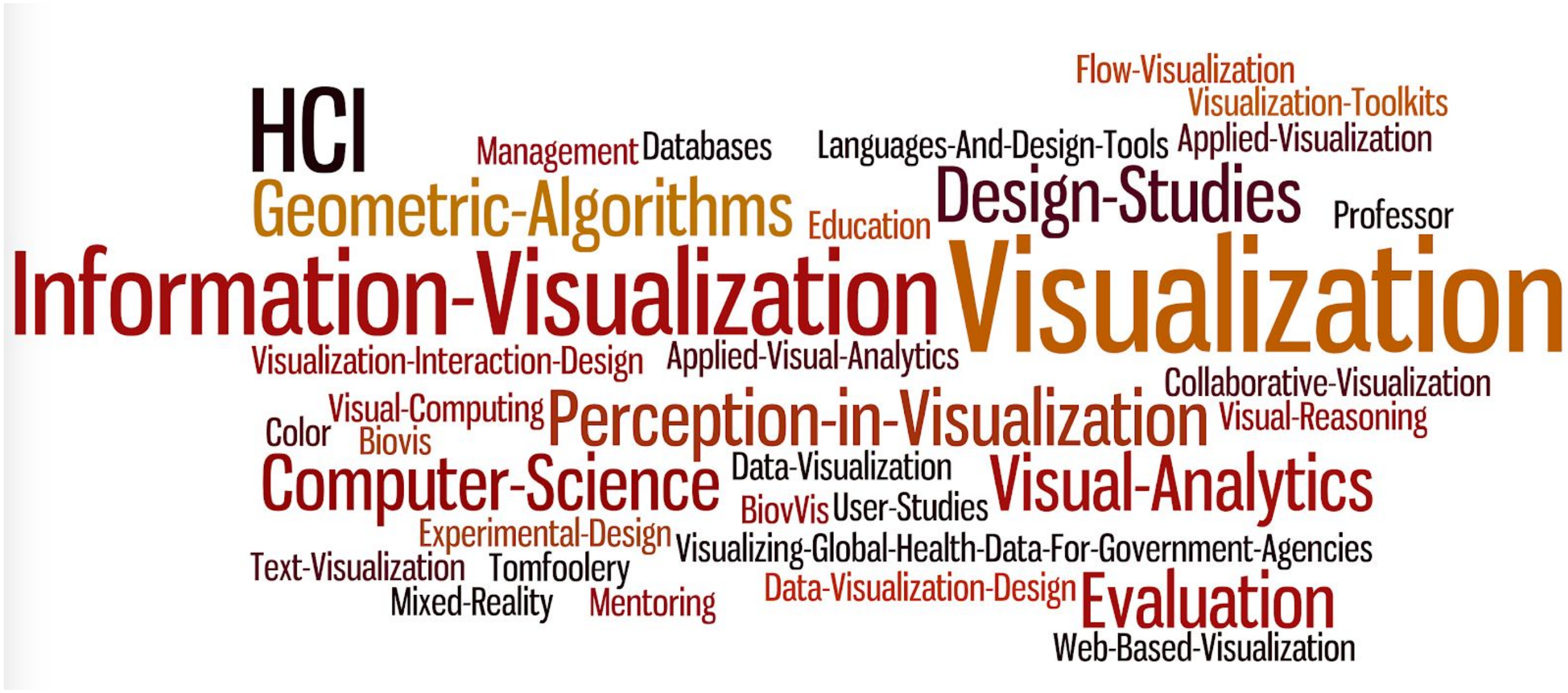

*Figure 1. Word cloud showing multi-term responses to the poll asking about expertise of panel attendees.*

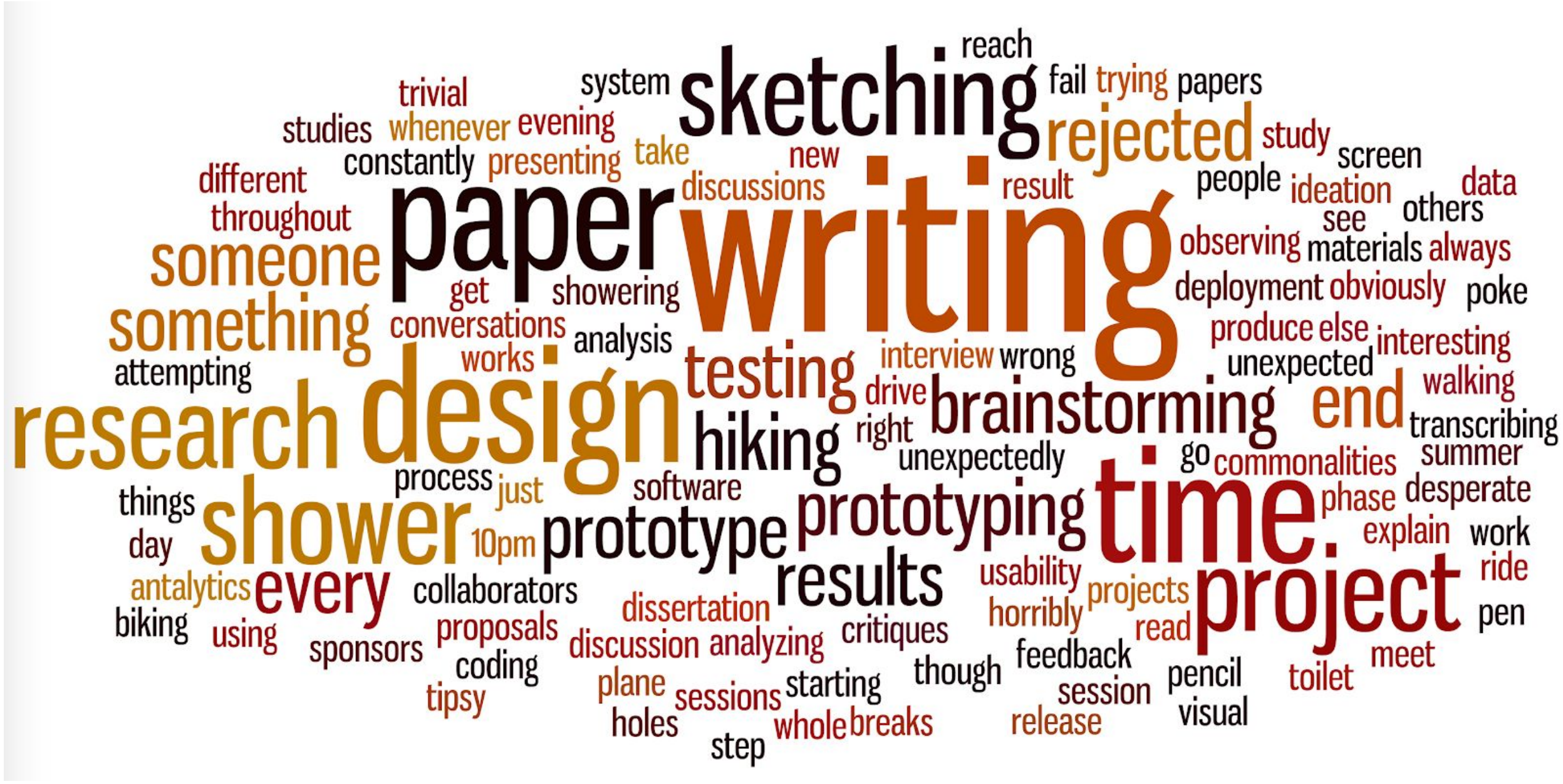

*Figure 2. Word cloud showing terms used in reporting when panel attendees reflected.*

The third poll question asked participants for ideas about reflection, and to specifically tag these ideas with a key letter to indicate if they were examples of best practice **[B]**, possible pitfalls **[P],** or open questions **[Q].** We assigned a key letter to any response without one, including assigning an **[X]** to three responses that did not easily fit into one of these categories. This poll was live during the question and answer period, and attendees were able to *up-vote* or *down-vote* responses that were displayed on screen.

.

The second data source is a summarized and anonymized transcript of the panel session: http://j.mp/googleDoc_IEEEVIS17_reflection. This summary captures the points made by both the panelists and the audience throughout the session, along with quotes that concisely (and sometimes colorfully) capture the sentiment of the responses. Points made by panelists are labeled P1-P6, audience points are labeled A1-A11, and organizer points are labeled ORGANIZER. The panel introduction by the organizers is not included in this summary, but is instead reflected in the introduction of this report.

# Key Themes

This section provides a high-level categorization of the content found in the panel data sources. Categories 1-4 explicitly consider ideas around reflection in design studies, and in visualization research more generally. Category 5 summarizes content focused broadly on the value of design studies, a theme that recurred in the question and answer phase and was initiated and developed by both audience and panelists.

**1. What to reflect on**

The importance of artifacts and insights to stimulate reflection was evident with several examples identified as being conducive to reflection. A range of different foci for the reflective process were suggested, including: visualizations as externalizations of mental models; moments of learning as pinpointed through questions by external researchers; and visualization design failures. Systems design research was given as an example beyond designs studies where reflection is critical.

**2. When to reflect**

A number of comments explicitly emphasized the role of paper writing at the end of a project as a clear opportunity for reflection. Others participants indicated that they reflect throughout the entire design process. It was noted that internal, personal reflection after the completion of a project is useful in identifying fruitful future directions.

**3. How to reflect**

Two specific modalities described in the comments for inducing reflection were writing and giving talks. Writing schemes include drafting a paper's abstract or introduction, developing a slide deck of ideas, or filling out a questionnaire early-on about a project's goals. Giving talks, particularly to external colleagues, was noted as an opportunity for prodding questions that reveal assumptions and internalized, learned knowledge. Open questions remain as to how to capture internalized and unintentional reflection, particularly over long periods of time and across multiple research projects.

**4. How to report reflections**

A significant number of comments focused on a lack of guidance or opportunity for reporting on reflection. Several pointed specifically to a need for more structured guidance on what goes into a reflection section of a research paper, including a richer description of the problem context. Reporting reflection on failures was seen as likely to provide valuable knowledge, but the lack of clear venue or mechanism for doing so was identified as a problem given the perceived importance of successful artifacts in academic papers.

**5. The value of design studies**

Many comments drifted from reflection as a process to instead address the value of design studies. Some noted the time-consuming nature of design study research, and questioned the trade-off of time with acquired knowledge. Others lamented the tension between the goals of visualization researchers, designers and other stakeholders, as as well as that between reporting design studies across broad audiences. As an indirect reference to reflection, several comments emphasized the value of design studies stemming from the learning that occurs during the design process, as opposed to the resulting software artifact.

# Priorities: Reflection on Reflection on Reflection

This report is not intended to be a research paper; our goal is to log data for the community and to informally summarize key ideas from what was an active and enjoyable panel session. Despite this, we feel compelled to make some suggestions based on what we learned for further thinking about how to use reflective practice to produce reliable knowledge in applied visualization research.

**Establish good practice for reflection**

Comments by both the panelists and the audience made it clear that many researchers in the visualization community actively engage with reflective practice, albeit in an ad-hoc and limited way.  We believe that a more structured and purposeful approach to reflection will enable the community to make better use of this practice, and to move the discipline forward in new and interesting ways.

While it is clear that reflection plays a role in how we synthesize months, or years, worth of work during the process of drafting an academic paper, there is little concurrence in the practice of structured reflection *during* the design process. When does reflection occur, what triggers it, and what should we be capturing from it? We speculate that this gap has led to lost insights and opportunities for learning, and more fundamentally, may pose a threat to the underlying validity of our design research.

As a first step, the community needs rich(er) descriptions of design activities and reflective synthesis. This can address some of the challenges associated with: developing evidence to support claims; reducing possible cognitive biases associated with memory at paper writing time; and linking evidence across applied visualization projects in meta-studies. But there are also clear difficulties and tensions with developing and using these rich descriptions — the processes for logging information and ongoing reflection must be useful, manageable, and not inhibit what are naturally rapid and reactive design processes; the processes for synthesizing this information must be reported and robust.

**Develop appropriate formats for recording and reporting reflections**

The lack of consensus on what is useful to include in rich description points to the need for varying levels of granularity in both the recording and reporting of reflective synthesis. While some panelists encouraged recording and reporting *everything* — from screenshots to transcripts to notes to slide decks — others pleaded for annotations and reflective summarizations to accompany the artifacts. We believe that multiple levels of granularity are important as different levels are useful for different types of analysis: low-level, raw artifacts and reflections may be a source of inspiration for transferring ideas to other problems and domains, while high-level reflective synthesis will benefit meta-analysis across multiple projects.

The familiar *overview+detail* approach for navigating complex data could be a useful model for organizing and guiding reflective practice. The ways that a rich description could be used — for the reflective researcher throughout the project, as inspiration for other projects, in validation of insights, for studying patterns across many projects — necessitate varying levels of details and analysis, as well as an accessible organizational structure. This requires a reporting outlet free from the constraints of a traditional academic paper; supplemental materials, design reports, and companion repositories offer opportunities that we may wish to explore.

**Understand the role of reflection in existing methodologies**

Despite the extensive use of reflective practice in the visualization community, and the intense and animated dialog in the panel session, we note a lack of discussion around the relationship between reflection and research methodologies, or the role of reflection in validating findings and conclusions. The explicit statement in the design study methodology that "*reflection is where research emerges from engineering*" [1] acknowledges the critical role of this practice in generating new knowledge — however, the community has yet to clearly define or deeply integrate reflection into the core methods and methodologies used by visualization researchers today. In our view, it is critical that we rectify this gap to ensure that the research findings that emerge from our engineering are credible and reliable.

**Continue the healthy discussion about design studies**

To our surprise — and despite our best efforts to focus on reflection — discussions throughout the panel session involved a clear and persistent focus on the nature of design studies, their value to the visualization research community, and the nuts-and-bolts of conducting them successfully. This debate was informative and useful, and a nice example of reflection in practice. It drew attention to some important issues associated with design study research: the risks involved due to their time-intensive commitment, and the problems with failure for young researchers and external collaborators. Even though design studies are now a common and well-defined approach to visualization research, there is still significant opportunity to further refine their role in what, and how, we study as visualization researchers. We speculate that

reflection — and our lack of understanding and consensus of this practice — underlies these continued debates.

# Conclusion

The *Reflection On Reflection In Design Studies* panel at the 2017 IEEE VIS conference provided plenty of evidence that reflection is occurring and valued in visualization design research. It confirmed our suspicions established through discussion of our own practices that the *when*, *how*, and *what* of reflection vary widely. The priorities in this report reflect our own interpretations of the themes that we took from the panel, as well as our own experiences of using reflection. We would like to thank the panelists and attendees for their insightful contributions to the debate, and the conference panel chairs for providing the platform. The panel helped us identify a significant opportunity for the community to define how we can use reflection effectively in our research approaches, how we can judge work that relies upon it in a consistent and fair way, and how we can use knowledge acquired through reflection to improve understanding of our domain.

We need to work out the role of reflection in visualization research and how we can use this practice effectively — learning from other disciplines could provide a useful first step. To start, we need to develop ideas about the key themes identified in this report: when to reflect, what to reflect on, and how to structure, report, use, and validate reflection within and beyond the context of an academic research paper. We believe that structured guidance for reflective practice has the potential to increase the reliability, quality, and impact of applied visualization research, and that the activity, themes and priorities that we report here move us forwards in community efforts to achieve this.